\documentclass[%
 aip,
 amsmath,amssymb,
 reprint,%
]{revtex4-1}

\usepackage{graphicx}
\usepackage{dcolumn}
\usepackage{bm}

\usepackage[utf8]{inputenc}
\usepackage[T1]{fontenc}
\usepackage{mathptmx}
\usepackage{float}
\begin{document}

\preprint{AIP/123-QED}

\title[]{Role of electron and ion irradiation in a reliable lift-off process with e-beam evaporation and a bilayer PMMA resist system}

\author{Bin Sun}
\email{sun@iht.rwth-aachen.de}
 \affiliation{ 
Institute of Semiconductor Electronics, RWTH Aachen University, 52056 Aachen, Germany
}%
 \author{Thomas Grap}
 \affiliation{ 
Institute of Semiconductor Electronics, RWTH Aachen University, 52056 Aachen, Germany}
\affiliation{%
Helmholtz Nano Facility of the Forschungszentrum Jülich, 52428 Jülich, Germany
}%
\author{Thorben Frahm}%
\affiliation{ 
Institute of Semiconductor Electronics, RWTH Aachen University, 52056 Aachen, Germany}
\author{Stefan Scholz}%
\affiliation{ 
Institute of Semiconductor Electronics, RWTH Aachen University, 52056 Aachen, Germany}
\author{Joachim Knoch}%
 \email{knoch@iht.rwth-aachen.de}
\affiliation{ 
Institute of Semiconductor Electronics, RWTH Aachen University, 52056 Aachen, Germany
}%

\date{\today}

\begin{abstract}
This paper addresses issues related to cracking and blisters in deposited films encountered in a lift-off process with electron beam evaporation and a bilayer PMMA resist system. The impact of charged particles, i.e. electrons and ions, is investigated using an electron beam evaporation chamber equipped with ring-magnets and a plate electrode placed in front of the sample. 
By replacing the plate electrode with a hollow cylinder, the modified evaporation setup utilizing passive components allows a complete elimination of resist shrinkage and blistering yielding near perfect deposition results for a large variety of different materials.
\end{abstract}

\maketitle

\section{\label{sec:Introduction}Introduction}
The lift-off process is a widespread and indispensable patterning technique for creating micro- and nano-structures of a target material on a substrate (e.g. silicon wafer) using a sacrificial layer (e.g. resist) \cite{enwiki:1001020615}. The technique is especially useful when there is no appropriate etching method that provides good selectivity between the target material and the substrate. 

The process of e-beam evaporation consists of placing the target material in a crucible and vaporizing the top surface in a vacuum chamber by supplying kinetic energy from an accelerated electron beam. The line-of-sight deposition nature is ensured by keeping the chamber at a high vacuum (\textasciitilde\,$10^{-6}$\,mbar) \cite{sarangan_nanofabrication_2019, chambers_basic_1998} such that collisions between evaporant material and background gas molecules are kept at a low level. 
In general, an inwardly tapered resist profile in combination with a line-of-sight deposition method, e.g. electron beam evaporation, is essential to facilitate solvent attack from resist sidewalls in order to obtain the best outcome in a lift-off process. 

Polymethyl methacrylate (PMMA) was one of the first materials developed for electron beam lithography and it still remains as the most commonly used positive-tone high-resolution e-beam resist. Moreover, an inwardly tapered resist profile can be easily realized by combining a more sensitive bottom layer (i.e. short polymer chain with lower molecular weight) and a less sensitive top layer (i.e. long polymer chain with higher molecular weight). As a result, a lift-off process utilizing e-beam evaporation in combination with bilayer PMMA is widely used especially in research labs, to create high-resolution metallic gate structures, for instance \cite{doh_tunable_2005,Huang2008,Bjork2009,brauns_palladium_2018,fathipour_exfoliated_2014}.
\begin{figure}[!h]
\centerline{\includegraphics[width=\columnwidth]{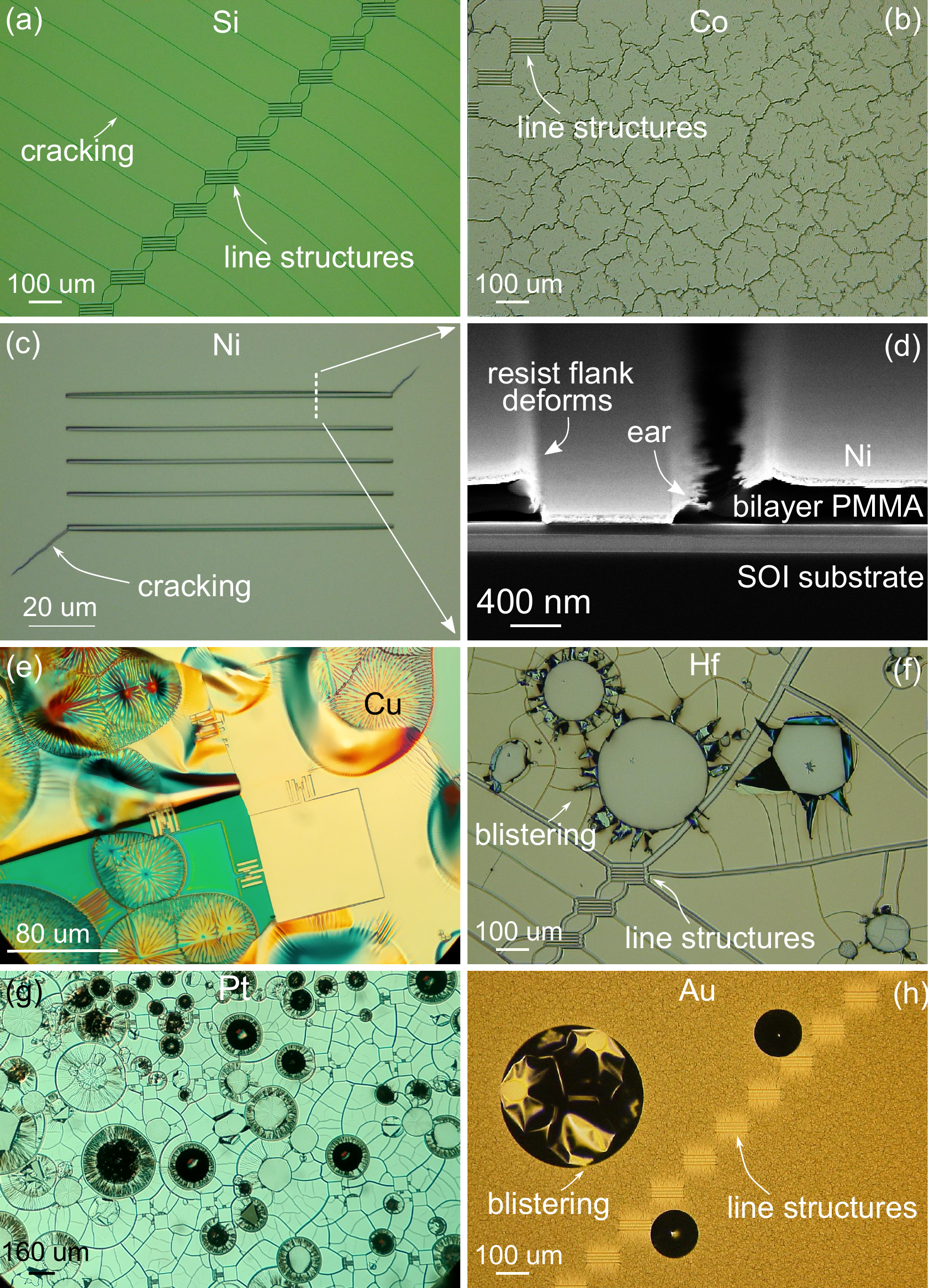}}
\caption{Typical results obtained with electron-beam evaporation. Optical microscopy image of Si (a), Co (b), Ni (c), Cu (e), Hf (f), Pt (g) and Au (f) on a bilayer PMMA system. A scanning electron micrograph cross-section along the dashed line in (c) is shown in (d). The sample surface is grounded by a metallic spring for (a)-(d), (f), (h) whereas the sample is glued on a tape, hence electrically floating for (c), (e). The substrate for (c), (d) is silicon-on-insulator whereas the rest is bulk silicon. Materials of low atomic number tend to cause cracking whereas blistering and cracking are often the case for high atomic number elements.}
\label{fig:bad_results}
\end{figure}
One of the major drawbacks associated with the lift-off process utilizing e-beam evaporation and PMMA resist system is related to resist shrinkage and bubble formation that lead to cracking and blistering of the deposited layer after the evaporation process\cite{Pao_solutionto}. Figure~\ref{fig:bad_results} displays typical results obtained for different evaporated materials (see section~\ref{sec:Experimental} for experimental details). While (a) - (c) depict optical microscopy images clearly showing cracking, an electron micrograph cross-section of nickel deposited onto a PMMA lift-off pattern is displayed in (d) revealing substantial shrinkage that leads to a strongly deformed resist pattern and hence the metal film. It is observed that materials of low atomic number often causes cracking induced by the shrinkage of PMMA films. High atomic number elements (approximately after Ni, i.e. Z\,\textgreater\,28) can lead to blistering as well as cracking as depicited in (e)-(h). 
Only very few studies on the subject have been published so far: Lei \textit{et al.} used a fluorine-based plasma to change the properties of PMMA \cite{Lei2020}. It is well known that PMMA will start to cross-link at a higher dose of electron irradiation (approximately 50 times the optimal dose for positive tone applications) \cite{Zailer_1996} and thus turns into a negative-tone resist. Accordingly, Pao \textit{et al.} suggested that the resist blistering stems from irradiation of the resist with electrons back-scattered from the crucible. They developed a screening apparatus using negatively biased plates to repel electrons that indeed mitigated blistering \cite{Pao_solutionto}. Very recently, Volmer \textit{et al.} demonstrated low energy electrons could also cause cross-linking and blistering by direct exposure under a filament. In addition, undesired shadow deposition is also mitigated with a negatively biased electrode to deflect ions \cite{volmer_how_2021}.

However, a thorough study investigating the dependence of the evaporation result on the evaporated material and its impact on the profile of the resist after e-beam deposition is missing. We employ an insulated electrode in front of the sample in order to measure the amount of charge hitting the sample during a typical deposition process. Our finding shows that electrons as well as positively charged ions play a role deteriorating the deposition results. Finally, we show that a combination of two passive measures - an appropriate magnetic field of two permanent magnets together with a cylinder-shaped, self-charging electrode - allows the complete elimination of resist shrinkage and blistering yielding near perfect deposition results for a large variety of different materials.
\section{\label{sec:Experimental}Experimental Set-up}
A high-vacuum electron beam evaporation tool (Balzers\textsuperscript{\textregistered} PLS 500) is used to evaporate different materials. The 270\,$^{\circ}$ e-beam source is operated at an acceleration voltage of 10\,kV. The distance between the crucible liner and substrate holder is \textasciitilde\,280\,mm. The shutter is located  \textasciitilde\,155\,mm above the crucible. Although the high-vacuum chamber is compact, we ruled out any impact of temperature (see appendix A) and X-rays (see below) on the PMMA. 

\begin{figure}[h]
\centerline{\includegraphics[width=\columnwidth]{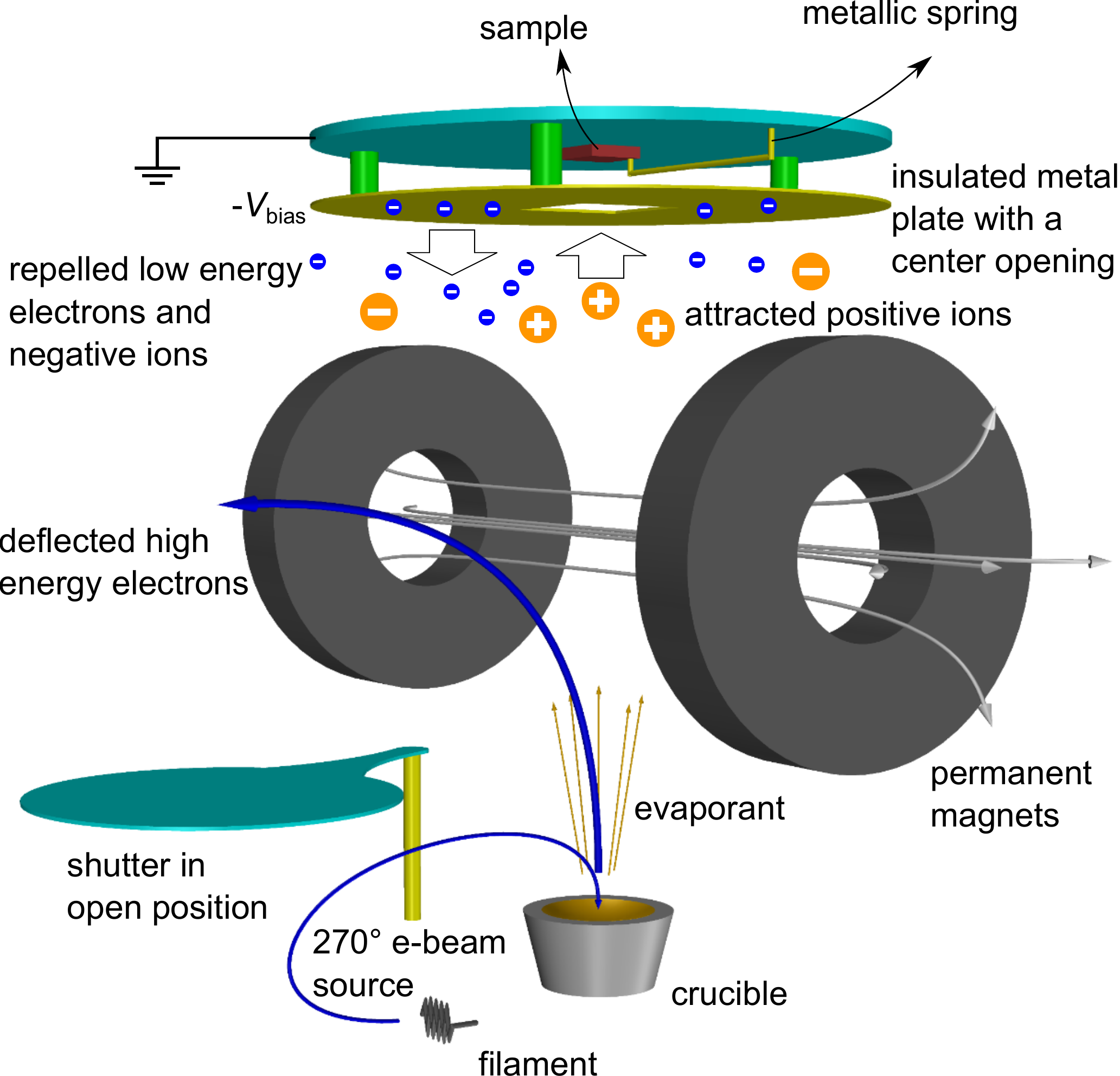}}
\caption{Schematic (dimension is not to scale) of a modified electron beam evaporation setup with permanent magnets and an insulated metal plate installed to avoid electron radiation on sample surface.}
\label{fig:sun1}
\end{figure}

Problems associated with resist blistering and shrinkage as well as unintentional shadow deposition are due to a bombardment of the PMMA with charged particles (electrons and ions \cite{Pao_solutionto,doi:10.1116/1.1315820,doi:10.1063/1.352334,volmer_how_2021}). A series of complex mechanisms underlie the generation of these electrons and ions. For simplicity, we distinguish between high and low energy particles only in the following. High energy electrons mainly stem from backscattered electrons reflected from the target material having energies on the order of 10\,\%-90\,\% of the initial beam energy \cite{doi:10.1116/1.1315820}. The low energy electrons mainly comprise secondary electrons, electrons created through thermionic emission, thermal ionization and electron-impact ionization, and photoelectrons, etc. These low energy electrons typically feature an energy up to \textasciitilde\,50\,eV \cite{volmer_how_2021, doi:10.1063/1.332840, doi:10.1063/1.352334}. Ions originate from thermal ionization and electron-impact ionization also feature a low energy typically up to several eV \cite{doi:10.1063/1.352334}. 

\begin{figure}[h]
\centerline{\includegraphics[width=\columnwidth]{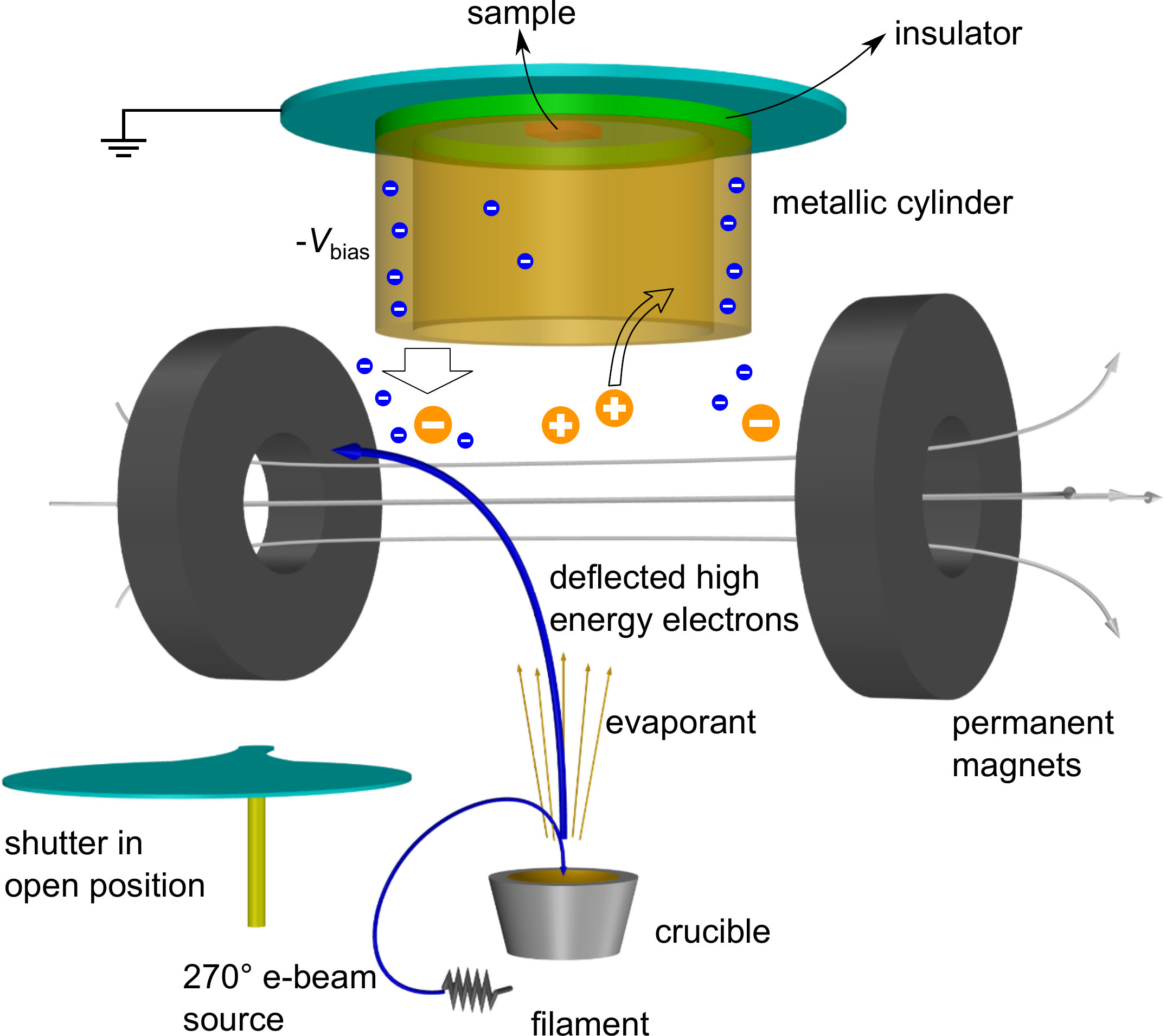}}
\caption{Schematic (dimension is not to scale) of the improved electron beam evaporation setup with permanent magnets and an insulated metallic cylinder installed to avoid electron radiation on sample surface.}
\label{fig:cylinder}
\end{figure}

Since charged particles can be manipulated with electric and magnetic fields, we equipped the evaporation tool with two permanent ferrite magnets\footnote{The inner diameter $d_\text{inner}$=60\,mm, the outer diameter $d_\text{outer}$=100\,mm, thickness 20\,mm. The magnets are installed relatively close to the substrate holder (bottom of the magnets to the substrate holder \textasciitilde\,125\,mm) to reduce the disturbance on the 270\,$^{\circ}$ e-beam source. The distance between the two magnets is \textasciitilde\,125\,mm.} and an additional electrode that is mounted in front of the sample but is insulated from the sample holder; this electrode is either realized as a simple plate or in the form of a cylindrical electrode. The two setups are schematically illustrated in Fig.~\ref{fig:sun1} and Fig.~\ref{fig:cylinder}. The additional electrode can either be used as an active electrode by applying a voltage, or it can also be used as a probe to measure the bombardment with charged particles. As it turns out (see discussion below), a major benefit of our setup is that for not too high deposition rates, a floating cylindrical electrode in combination with permanent magnets completely eliminates resist shrinkage and blistering. Hence, a passive setup that neither requires any additional installation nor feedthroughs into the high vacuum chamber is effective. 

The plate electrode, as illustrated in Fig.~\ref{fig:sun1}, is placed \textasciitilde\,10\,mm below the substrate holder. It is \textasciitilde\,125\,mm in diameter and the opening at the center is 25\,mm$\times$20\,mm in size. In the case of the cylindrical electrode (depicted in Fig.~\ref{fig:cylinder}), it is directly glued on the substrate holder with a piece of glass in between. The cylindrical electrode is 25\,mm in diameter and 40\,mm in depth (aspect ratio 1.6). As mentioned above, both electrodes can either be operated in passive mode or actively biased with an external voltage source (with DC voltages in the range -650\,V to +650\,V) via two cable feedthroughs. In the former case, a Fluke\textsuperscript{\textregistered} 179 multimeter is used to measure the potential accumulated on the electrode during an evaporation process. Potential measurements are carried out for a variety of materials including metals, semiconductor and insulators. 
The choice of materials encompasses typical materials in nanotechnology covering a broad range of atomic number and  melting point.

\section{\label{sec:Results}Impact of Charged Particles}
When carrying out an electron beam evaporation the beam current is first increased with the sample protected by a shutter until a certain deposition rate is reached. One could therefore argue that a bombardment with electrons occurs only during the rather short period where the shutter is opened and the deposited (metallic) film is not yet thick enough and hence a high deposition rate would be preferable.

To study the impact of a bombardment of a PMMA resist layer with electrons, bulk silicon substrates (\textit{p}-type boron doped, 5-10\,$\Omega\cdot$cm) are prepared with spin coating of bilayer PMMA films (bottom layer AR-P\,639 and top layer AR-P\,679 from Allresist GmbH). Line structures (1\,$\mu$m in width) are patterned with electron beam lithography and development in a mixture of methyl isobutyl ketone (MIBK) and isopropanol (IPA) (MIBK:IPA=1:3). 
The samples are mounted into the e-beam evaporation chamber (without magnets and electrode) and the emission current is increased up to 35\,mA using a Pt target; note that at 35\,mA beam current, no evaporation rate is detected with a quartz crystal microbalance and thus, a direct irradiation of charged Pt ions, neutral Pt atoms and high energy electrons on PMMA film can be excluded with a closed shutter.

\begin{figure}[H]
\centerline{\includegraphics[width=\columnwidth]{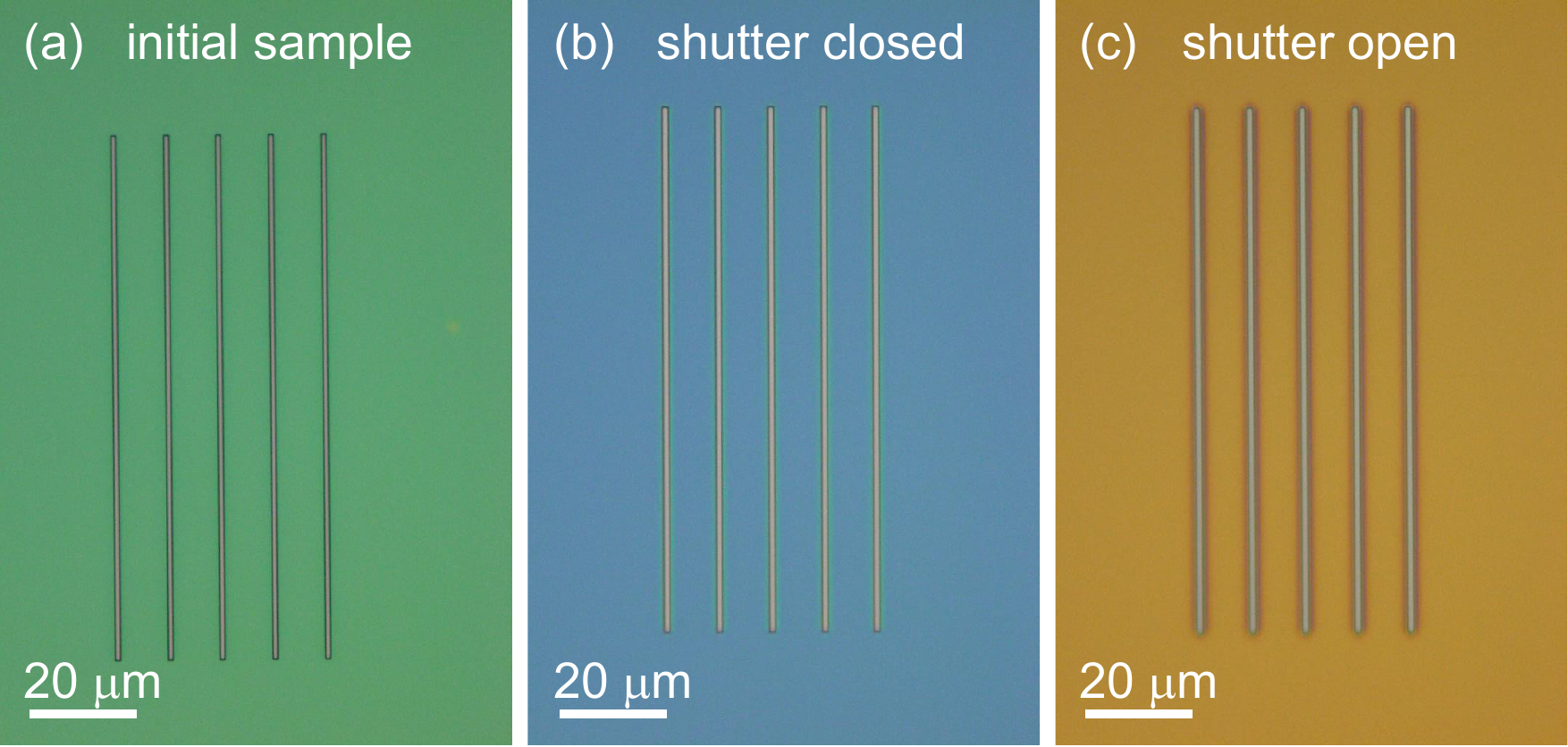}}
\caption{Comparison of bilayer PMMA film color for different evaporation conditions using platinum. All samples initially exhibit a green color after PMMA coating, electron beam lithography, and development like the reference sample (a). Film color changed to blue after ramping up power to 35\,mA while keeping the shutter closed for 15\,min (b). Film color changed to braun after ramping up current to 35\,mA and open shutter for 15\,min (c).}
\label{fig:film}
\end{figure}
For a rapid initial study, optical microscopy of the PMMA film is employed using a Keyence\textsuperscript{\textregistered} digital microscope. All samples initially exhibit a green color as displayed Fig.~\ref{fig:film}(a). The PMMA film color changes into blue (see Fig.~\ref{fig:film} (b)) after ramping up the beam current to 35\,mA while holding the shutter closed for 15\,min (shutter is also kept closed for beam current ramping up/down). This is unexpected and means that a substantial fraction of the electron beam incident on the target reaches the sample via multiple backscattering from the target and chamber walls. In addition, high energy electrons hitting the chamber walls may also lead to secondary electron emission that is incident on the sample. Next, when the shutter is opened for 15\,min after reaching 35\,mA beam current while keeping all other parameters the same, the PMMA film color changes into brown (as shown in Fig.~\ref{fig:film}(c)). Thus, we conclude that high energy electrons have a strong impact on the PMMA even during the beam current ramp-up when the shutter is closed.  

\begin{figure}[h]
\centerline{\includegraphics[width=\columnwidth]{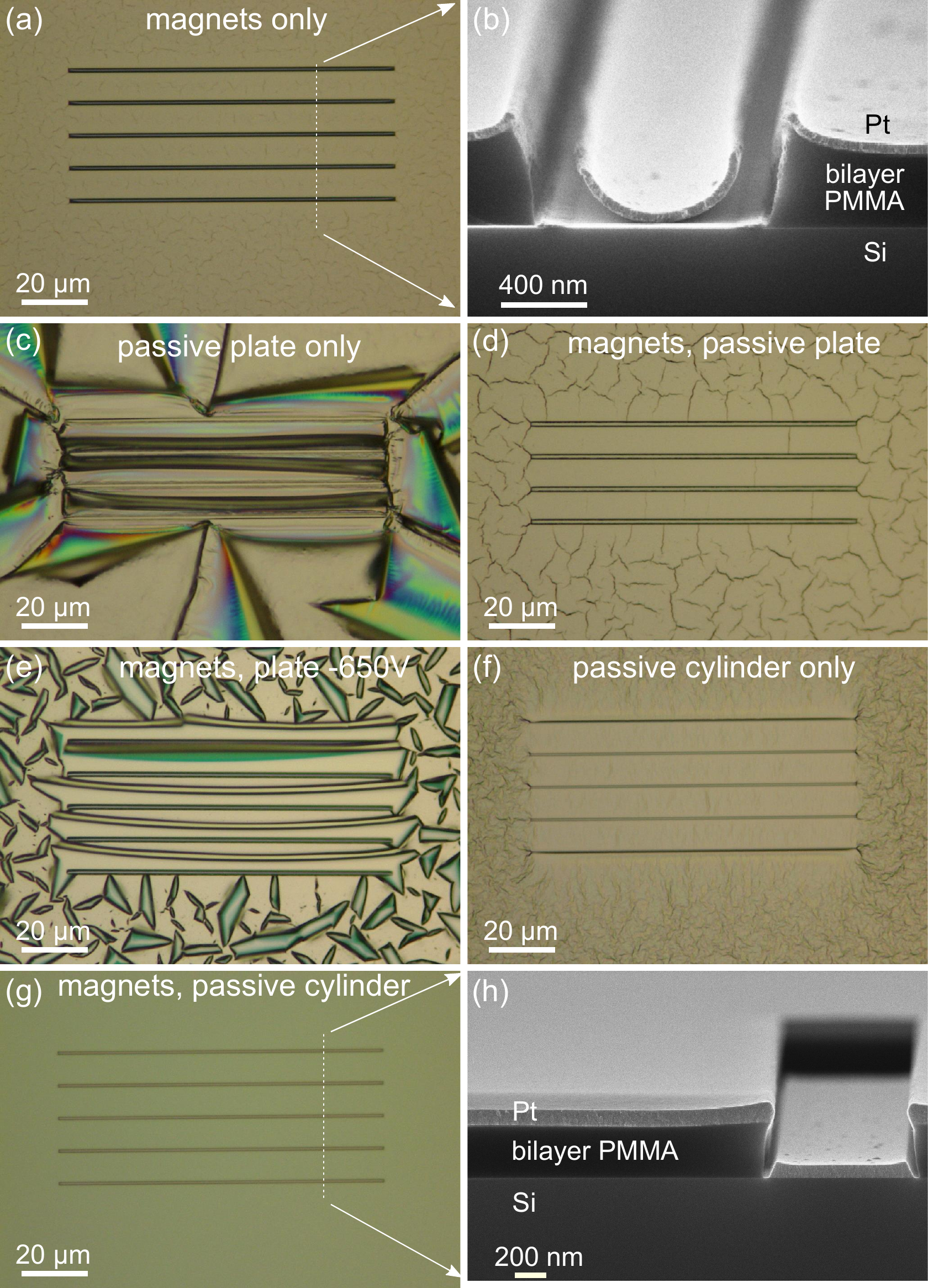}}
\caption{Passive plate only (a). Magnets only (b). Magnets and passive plate (c). Magnet and the plate is actively biased at -650\,V with an external DC power supply (d). Magnet and the plate is actively biased at +650\,V (e). Passive cylinder only (f). Magnet and passive cylinder (g) and a cross-sectional scanning electron micrograph along the dashed line in (g) is shown in (h). Recorded potential and other evaporation parameters for (g) is plotted in (i).}
\label{fig:Pt_mix}
\end{figure}

Since charged particles can be manipulated with electric and magnetic fields, the two ferrite ring magnets together with the plate or the cylindrical electrode depicted in Fig.~\ref{fig:sun1} and \ref{fig:cylinder} are mounted in different configurations and their impact on the deposition result is studied. To this end, \textasciitilde\,80\,nm of platinum is evaporated with a beam current of \textasciitilde\,81\,mA (static spot \textasciitilde\,3\,mm in diameter centered at the crucible) yielding a deposition rate of 1\,\AA/s. The film surface depicted in Fig.~\ref{fig:Pt_mix}(a) shows slight cracking if only the magnets are installed showing that the magnets are very effective in deflecting the high energy backscattered electrons (compared with Fig.~\ref{fig:bad_results}(g)). However, despite the absence of blistering, the cross-sectional scanning electron micrograph shown in Fig.~\ref{fig:Pt_mix}(b) reveals a bended Pt film (particularly on top of Si) and the layer starts to peel off the substrate. The reason for the peeling remains unclear but appears to be due to Pt ions involved in the deposition process leading to reduced adhesion of the Pt film on the substrate. 

Next, the magnets are removed and an insulated plate electrode is installed. If it was only electron bombardment that yields blistering and resist shrinkage the plate should also be effective since it is expected that the plate negatively charges up (which is indeed the case as discussed further below) thus deflecting at least the low energy part of the electrons. However, as depicted in Fig.~\ref{fig:Pt_mix}(c), strong blistering is observed which is attributed to a significant bombardment of the sample with positively charged Pt ions that are accelerated by the negatively charged plate electrode. This is confirmed when mounting the magnets and the plate electrode showing slightly worse results compared to the case with magnets alone (compare Fig.~\ref{fig:Pt_mix}(d) with (a)). Furthermore, applying a large negative bias of -650\,V at the plate electrode results in even stronger blistering due to accelerated positive Pt ions as shown in Fig.~\ref{fig:Pt_mix}(e). The experiment continues by replacing the plate electrode with a cylindrical electrode as illustrated in Fig.~\ref{fig:cylinder} (however, no magnets are installed in the present case). As displayed in Fig.~\ref{fig:Pt_mix}(f) a significant improvement compared to the plate electrode alone (Fig.~\ref{fig:Pt_mix}(c)) is obtained showing that the cylindrical geometry of the electrode shields the sample from a part of the Pt ions. Finally, if the magnets and a floating cylindrical electrode are used, an ideal evaporation result could be achieved as depicted in Fig.~\ref{fig:Pt_mix}(g) and (h): No blistering is observed and a cross-sectional scanning electron micrograph of the line structure shows no resist shrinkage and no bending of the deposited Pt film.

\begin{figure}[h]
\centerline{\includegraphics[width=\columnwidth]{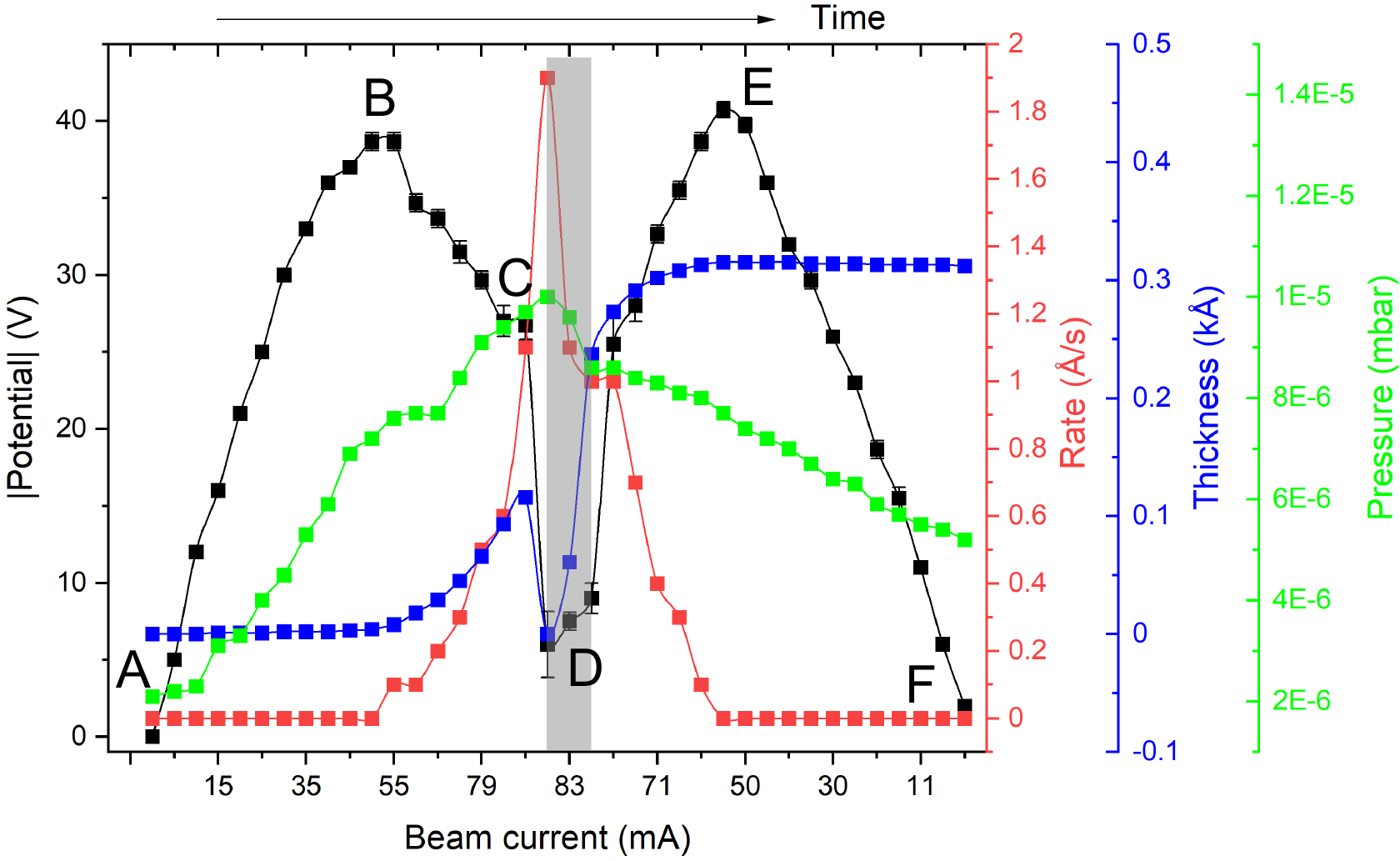}}
\caption{Potential at the cylindrical electrode, evaporation rate, film thickness (not corrected by tooling factor) and chamber pressure as a function of the electron beam current during Pt evaporation.}
\label{fig:Pt_plate}
\end{figure}

To further elaborate on this, the cylindrical electrode is used to measure the potential of the electrode due to the charge accumulated during an electron beam deposition process. This potential together with the evaporation rate, deposited layer thickness and chamber pressure are plotted in Fig.~\ref{fig:Pt_plate} as a function of the electron beam current (note, that the $x$-axis can also be interpreted as a time axis). First, the potential increases from A to B with an increasing beam current before evaporation becomes non-zero reconfirming the observation shown in Fig.~\ref{fig:film} (b). When increasing the electron beam current from \textasciitilde\,55\,mA to \textasciitilde\,81\,mA, the potential gradually drops while the evaporation rate increases (B to C). It is important to note that in this stage of the deposition process the shutter is still closed and hence the potential drop is likely caused by secondary electron emission due to a bombardment with high energy electrons reflected from the chamber walls. In addition, more material in the crucible turns from solid into liquid phase and an increased ionization at higher evaporation rate \cite{volmer_how_2021} might also account for the potential drop.
After the desired evaporation rate is reached, a sudden drop in plate potential is observed from C to D due to Pt$^{+}$ ions reaching the cylindrical electrode after opening the shutter (gray shaded area in Fig.~\ref{fig:Pt_plate}). For platinum the measured potential drops to nearly zero at an evaporation rate of \textasciitilde\,3\,\AA/s. When the shutter is closed, Pt ions are prevented from reaching the electrode and the potential increases accordingly and reaches approximately -40\,V again when the electron beam current is reduced; the potential is therefore \lq M\rq -shaped. 

Similar measurements of the potential during a deposition process have been carried out for Al$_2$O$_3$, Si, TiO$_2$, Co, Ni, Hf and Au, i.e. typical materials used in microfabrication with increasing atomic number and different melting temperatures (see appendix B). All materials show an \lq M\rq-shaped behavior of the potential. Moreover, the maximum potential measured is found to be strongly correlated with the atomic number of the material due to an increased backscattering of the electron beam from the target material. 
For some of the materials, e.g. Si, Hf, TiO$_\text{2}$, and Al$_\text{2}$O$_\text{3}$, the measured potential further increases or reaches a plateau with increasing evaporation rate. This can be explained with different ionization degrees for different materials. Different materials exhibit distinct properties, however, for most of the tested materials, a potential drop is observed during the switching of the shutter after reaching the desired evaporation rate (1\,\AA/s).
\begin{figure}[h]
\centerline{\includegraphics[width=0.6\columnwidth]{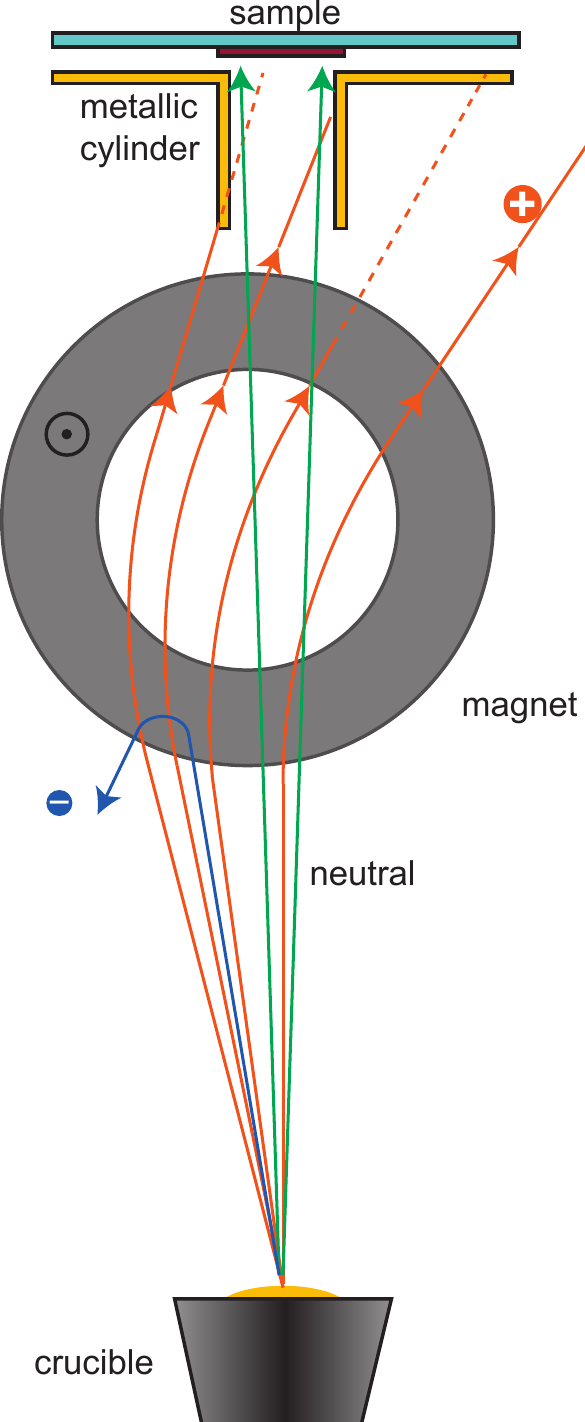}}
\caption{Geometry of the combined magnet/cylindrical electrode setup. The bending of the path of ions (orange) allows shielding of ion bombardment with the cylindrical electrode provided a suitable aspect ratio is chosen. The much lighter electrons are deflected by the magnetic field alone (blue). Deposition occurs only due to neutral Pt (green).}
\label{fig:Pt_coil}
\end{figure}

The fact that the combination of magnets with cylindrical electrode completely removes blistering and resist shrinkage even though the electrode potential drops to approximately zero during Pt evaporation can be understood based on the geometry alone. Figure~\ref{fig:Pt_coil} shows the movement of positively charged Pt ions (orange lines). Due to the Lorentz force the path of the ions is bent. This means that in the case of a cylindrical electrode with sufficient aspect ratio the ions are unable to reach the sample. Moreover, since the electrons are much lighter than the ions they are effectively bent away (in opposite direction, blue lines) even in the case of high energy electrons and also do not bombard the sample anymore. As a result, a perfect deposition due to neutral Pt vapor (green lines) is obtained. 

\begin{figure}[H]
\centerline{\includegraphics[width=\columnwidth]{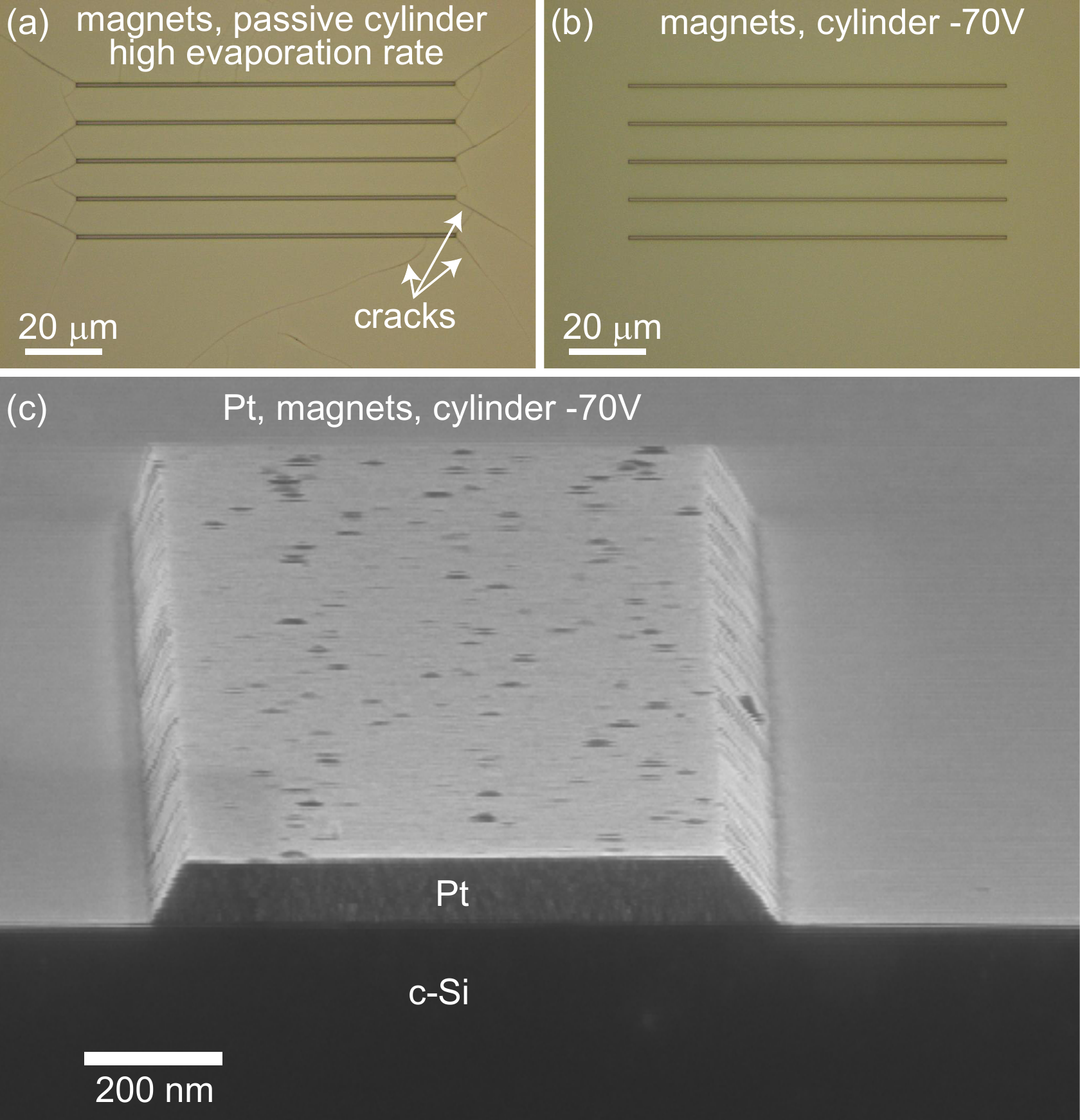}}
\caption{Magnet and passive cylinder with a higher evaporation rate of \textasciitilde\,3.8\,\AA/s (a). Magnet and active cylinder biased at -70\,V (b). Cross-sectional scanning electron micrograph of the sample shown in (b) after lift-off. Note that the Pt film adheres well to the silicon substrate without a Ti adhesion interlayer.}
\label{fig:Pt_liftoff}
\end{figure}

Interestingly, if the magnetic field strength and the aspect ratio of the cylindrical electrode remain unchanged but the deposition rate is increased, the potential accumulated on the electrode decreases and the velocity of the ions also increases. As a result, the bending and shielding action of the magnet and cylindrical electrode becomes less effective. Indeed, increasing the evaporation rate to \textasciitilde\,3.8\,\AA/s leads again to cracks in the resist (cf. Fig.~\ref{fig:Pt_liftoff}(a)). Three measures can then be taken to mitigate this cracking: i) an increase of the magnetic field strength, ii) an increased aspect ratio of the cylindrical electrode or iii) applying a voltage to the cylindrical electrode. In principle, both, positive and negative voltages should lead to additional bending/deflection of the positive ions. A positive voltage, however, has the risk of accelerating low energy electrons. Therefore, we applied a moderate negative bias of -70\,V, i.e. of the same order as the self-charging of the electrode during beam ramp-up. As displayed in Fig.~\ref{fig:Pt_liftoff}(b), this measure is in fact effective and a perfect lift-off result is obtained (cf. Fig.~\ref{fig:Pt_liftoff}(c)). The lift-off is easily accomplished in boiling acetone within 1\,min without mechanical agitation. This indicates that X-rays, which can not be manipulated by electric or magnetic fields, do not cause heavy crosslinking of PMMA throughout the evaporation process. Furthermore, it is important to note that the Pt film sticks well to the silicon substrate without any adhesion interlayer (such as Ti) which is usually necessary \cite{ababneh_electrical_2017}.\\
\indent In conclusion, the modified evaporation setup with appropriate magnets and cylindrical electrode with adequate aspect ratio delivers an ideal evaporation scheme free of adhesion issues and complete elimination of resist blistering and shrinkage provided a moderate evaporation rate (< \textasciitilde\,1.5\,\AA/s)) is used. If a higher evaporation rate is desired, the cylindrical electrode has to be biased sufficiently with an external voltage supply. 

\section{Evaporation of other materials}
\begin{figure}[H]
\centerline{\includegraphics[width=\columnwidth]{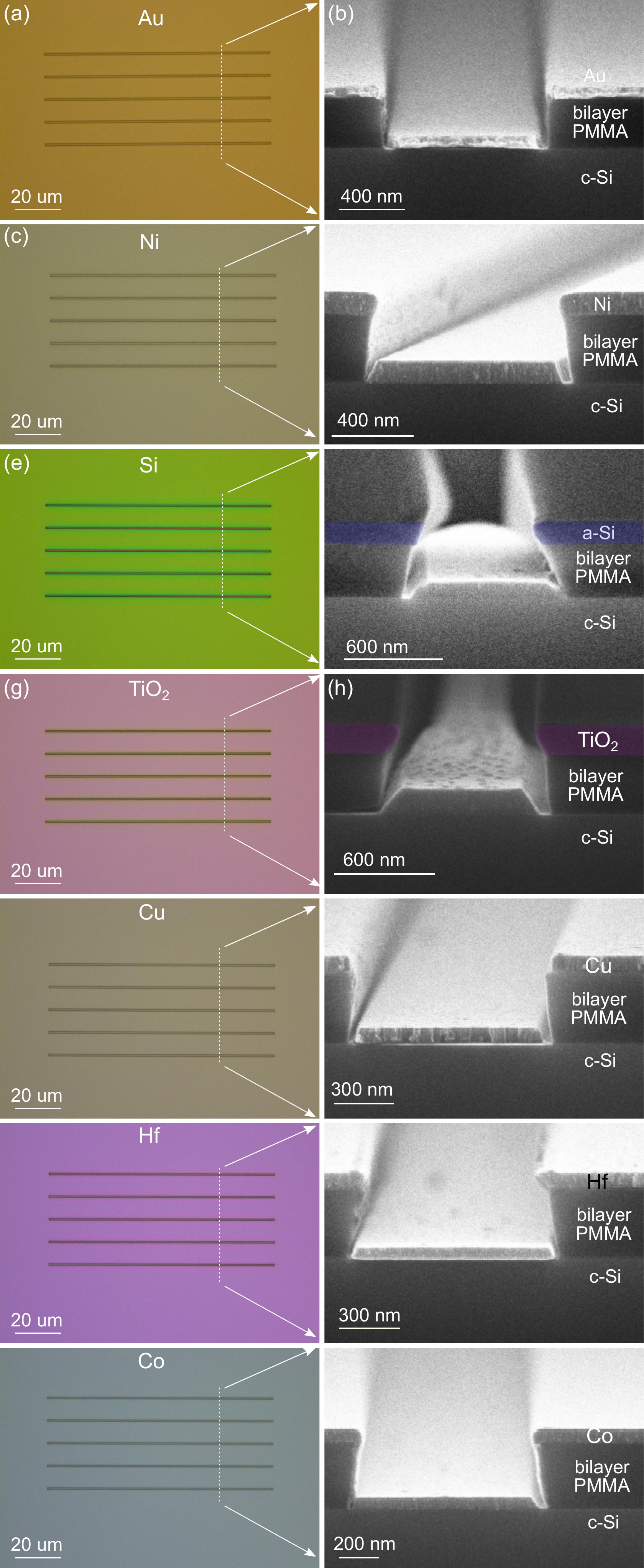}}
\caption{Evaporation of some other commonly used materials in semiconductor technology using the modified setup with magnets and a passive cylindrical electrode for Au (a), Ni (b), amorphous Si (c) and TiO$_\text{2}$ (d).}
\label{other_materials}
\end{figure}

Figure~\ref{other_materials} shows the evaporation results for a variety of materials using the magnet/cylindrical electrode-setup depicted in Fig.~\ref{fig:cylinder} at a moderate evaporation rate of 1\,\AA/s. It is clear that the resist profile remains intact after the evaporation, which facilitates solvent attack from the undeposited inwardly tapered sidewalls. 
Interestingly, the resist sidewalls are slightly covered with material in the case of Co (stronger) and Ni (less) which both are ferromagnetic materials. We speculate that this behavior is due to the fact that the magnets are installed relatively close to the substrate holder. The issue could be mitigated by placing the magnets further away from the substrate holder or by using a cylinder with a larger aspect ratio.

In conclusion, despite the difference in material properties, the modified evaporation setup equipped with magnets and the passive cylindrical electrode at a moderate evaporation rate (1\,\AA/s) is very effective in avoiding any irradiation with charged particles, yielding an almost ideal evaporation scheme for a variety of materials commonly used in semiconductor technology.


\section{\label{sec:summary}Summary}
We proposed a modified evaporation setup equipped with two magnets and a hollow cylindrical electrode. The passive components turn out to be very effective in avoiding hazardous bombardment of electrons and ions on PMMA. X-rays and heat are found to have a less impact. As it turns out, the modified evaporation setup allows a complete elimination of resist shrinkage and blistering yielding near perfect deposition results for a large variety of different materials. Furthermore, the proposed setup may also alleviate damages to the deposited layer itself which were speculated to be caused by X-rays and secondary ions\cite{doi:10.1116/1.3610989}.


\begin{acknowledgments}
Bin Sun is grateful for financial support from China Scholarship Council. The authors thank C. Teichrib (I. Institute of Physics A, RWTH Aachen University) for carrying out X-ray exposure experiments as confirmation. This work was partially supported by Deutsche Forschungsgemeinschaft (DFG, German Research Foundation) under Germany’s Excellence Strategy - Cluster of Excellence Matter and Light for Quantum Computing (ML4Q) EXC 2004/1 - 390534769, under KN545/22-1 and KN545/29-1.
\end{acknowledgments}

\appendix
\section{\label{subsec:temperature}Impact of heat}
To investigate the impact of heat, irreversible temperature test strips (RS PRO\textsuperscript{\textregistered}) are used to measure the highest temperature reached during the entire evaporation process through a color change of the sensing material. The test strips are mounted on a sample holder with the color change material facing towards the crucible. The test strips offer a measurement range between 40\,$^{\circ}$C and 260\,$^{\circ}$C.
\begin{table}[!h]
\begin{tabular}{ccccccccl}
\cline{1-8}
\multicolumn{1}{|c|}{}             & \multicolumn{1}{c|}{Au} & \multicolumn{1}{c|}{Pt} & \multicolumn{1}{c|}{Ni} & \multicolumn{1}{c|}{Co} & \multicolumn{1}{c|}{Hf} &\multicolumn{1}{c|}{Si} &\multicolumn{1}{c|}{Cu}\\ \cline{1-8}
\multicolumn{1}{|c|}{max beam current [mA]} & \multicolumn{1}{c|}{176}  & \multicolumn{1}{c|}{78}  & \multicolumn{1}{c|}{51}  &  \multicolumn{1}{c|}{49} & \multicolumn{1}{c|}{82} &\multicolumn{1}{c|}{49} &\multicolumn{1}{c|}{28}\\ \cline{1-8}
\multicolumn{1}{|c|}{Temperature [$^{\circ}$C]}  & \multicolumn{1}{c|}{< 99}  & \multicolumn{1}{c|}{<\,116}  & \multicolumn{1}{c|}{<\,71}  &  \multicolumn{1}{c|}{<\,71} &\multicolumn{1}{c|}{<\,110} &\multicolumn{1}{c|}{<\,65} &\multicolumn{1}{c|}{<\,60}\\ \cline{1-8}
\end{tabular}
\caption{The evaporation parameters of different metals are kept at the same initial pressure, evaporation rate (1\,\AA/s), and total thickness (80\,nm).}
\label{table1}
\end{table}\\
\indent Table~\ref{table1} depicts the maximum beam current used to achieve 1\,\AA/s evaporation rate and the maximum temperature reached during the entire evaporation process to deposit \textasciitilde\,80\,nm of materials that often suffer from peeling and blistering issues. It is clear that a higher beam current does not necessarily lead to a high temperature. According to the manufacturer of PMMA resist, the glass temperature is in the range of 105\,$^{\circ}$C, and the polymers are thermostable up to a temperature of 230\,$^{\circ}$C. The highest measured temperature is well below the Curie temperature of commercial ferrite magnets \cite{doi:10.1119/1.5124290, magnet_standard}.\\
\indent In conclusion, although the actual temperature seen by the PMMA film might deviate limited by the measurement method used, the measured temperature is very close to the glass temperature of PMMA. Therefore, it is recommended to water cool the substrate holder especially for deep submicron structures. Furthermore, permanent ferrite magnets do not lose magnetism throughout the evaporation process although some refractory metals are not tested, for instance tungsten and tantalum.

\section{\label{potential_other_materials}Potential measured on a cylindrical electrode with magnets installed}
Potential measured on a passive cylindrical electrode for some other materials. Magnets are also used in this measurement setup as indicated in Fig.~\ref{fig:cylinder}. It is observed that the measured potential value is heavily dependent on the beam spot position and whether wobbling is used. Therefore, the potential measurements are carried out with a static spot centered at the crucible regardless of the material used. As it turns out, almost all materials exhibit a `M'-shaped potential profile like the case for platinum as depicited in Fig.~\ref{fig:Pt_plate}.
\begin{figure}[H]
\centerline{\includegraphics[width=\columnwidth]{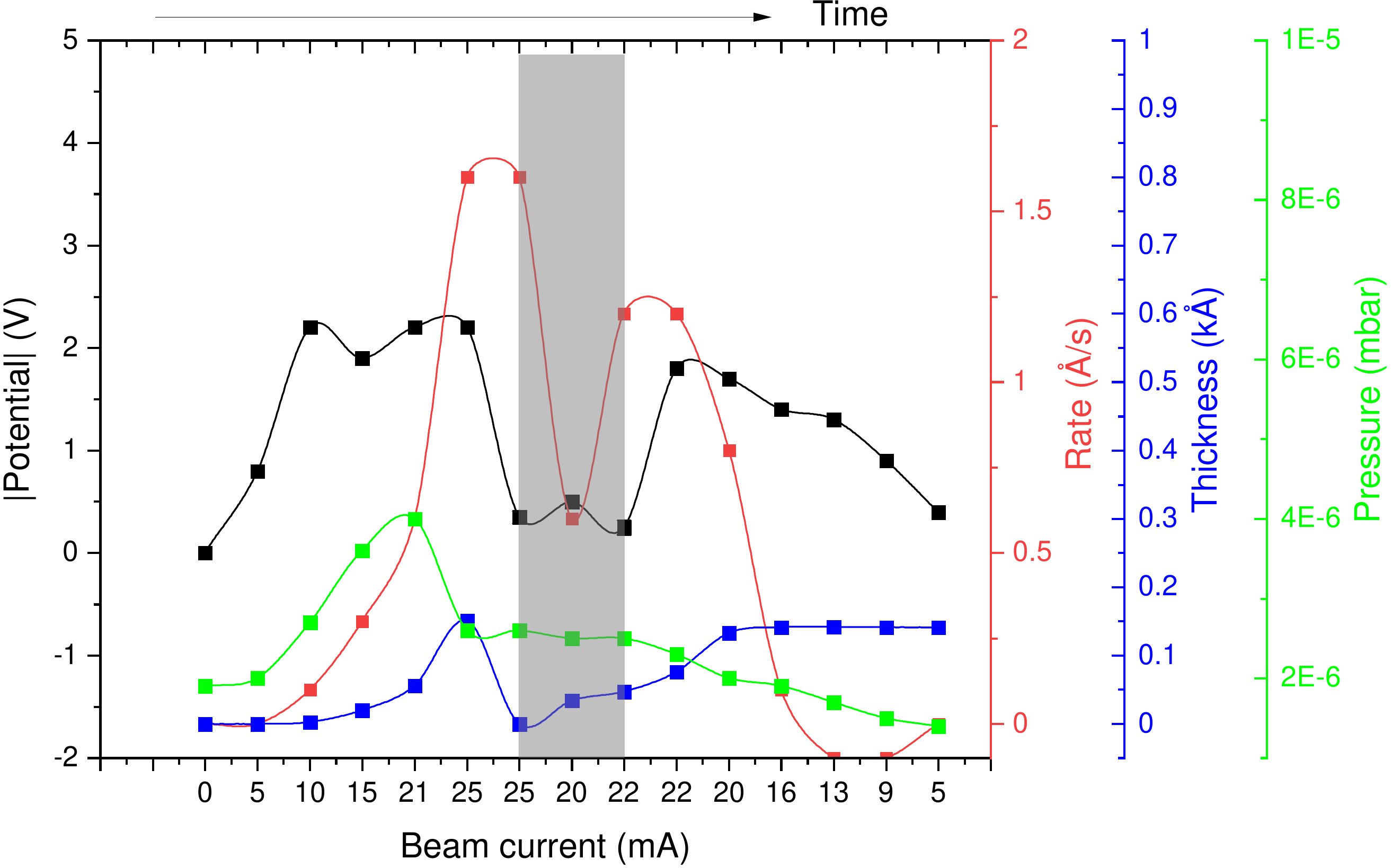}}
\caption{Potential at the cylindrical electrode, evaporation rate, film thickness (not corrected by tooling factor) and chamber pressure as a function of the electron beam current during Al$_\text{2}$O$_\text{3}$ evaporation. Time window for an open shutter is indicated in the shaded area.}
\label{fig:Al2O3_cylinder_magnet}
\end{figure}
\begin{figure}[H]
\centerline{\includegraphics[width=\columnwidth]{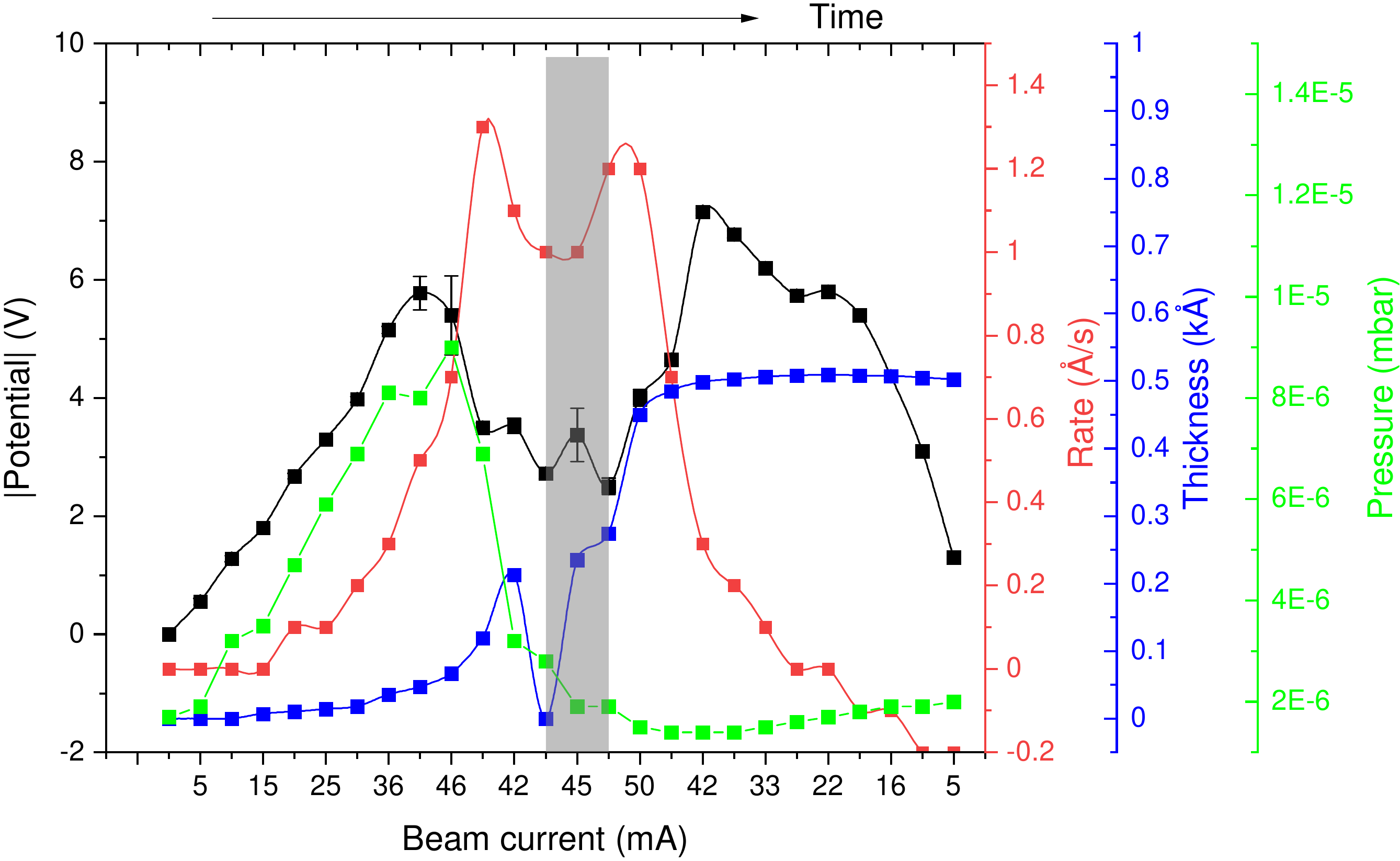}}
\caption{Potential at the cylindrical electrode, evaporation rate, film thickness (not corrected by tooling factor) and chamber pressure as a function of the electron beam current during Si evaporation. Time window for an open shutter is indicated in the shaded area.}
\label{fig:Si_cylinder_magnet}
\end{figure}
\begin{figure}[H]
\centerline{\includegraphics[width=\columnwidth]{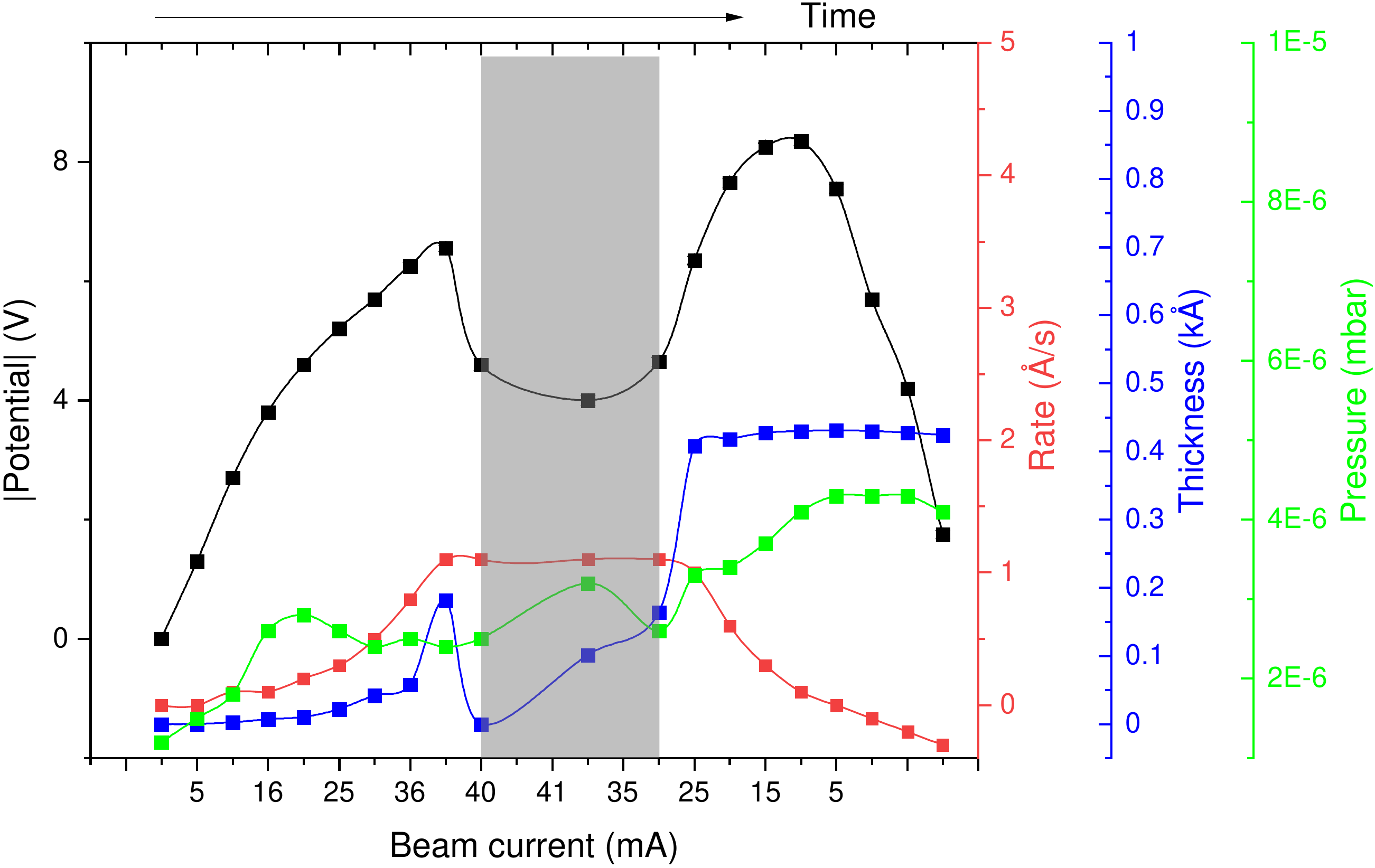}}
\caption{Potential at the cylindrical electrode, evaporation rate, film thickness (not corrected by tooling factor) and chamber pressure as a function of the electron beam current during TiO$_\text{2}$ evaporation. Time window for an open shutter is indicated in the shaded area.}
\label{fig:TiO2_cylinder_magnet}
\end{figure}
\begin{figure}[H]
\centerline{\includegraphics[width=\columnwidth]{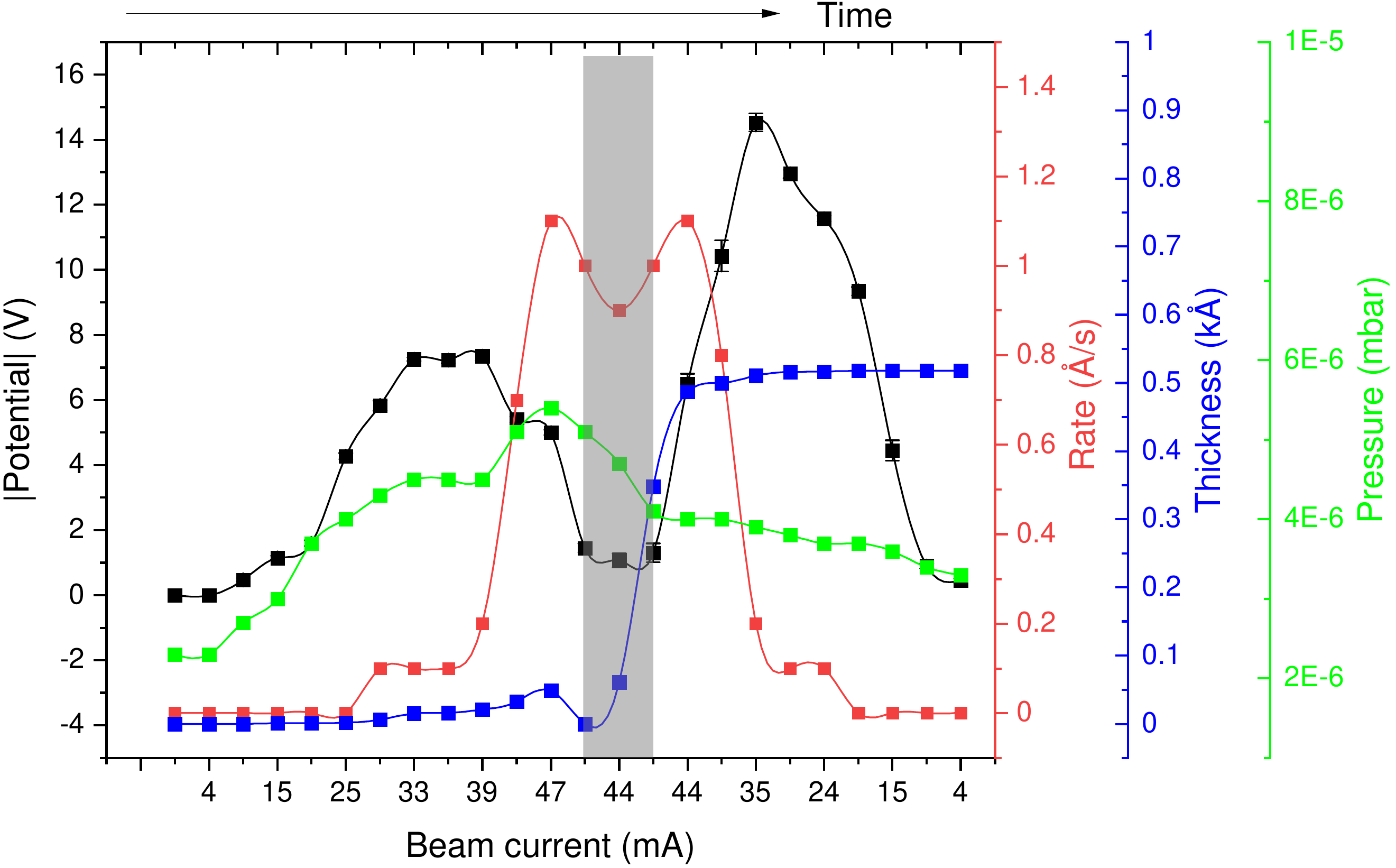}}
\caption{Potential at the cylindrical electrode, evaporation rate, film thickness (not corrected by tooling factor) and chamber pressure as a function of the electron beam current during Co evaporation. Time window for an open shutter is indicated in the shaded area.}
\label{fig:Co_cylinder_magnet}
\end{figure}
\begin{figure}[H]
\centerline{\includegraphics[width=\columnwidth]{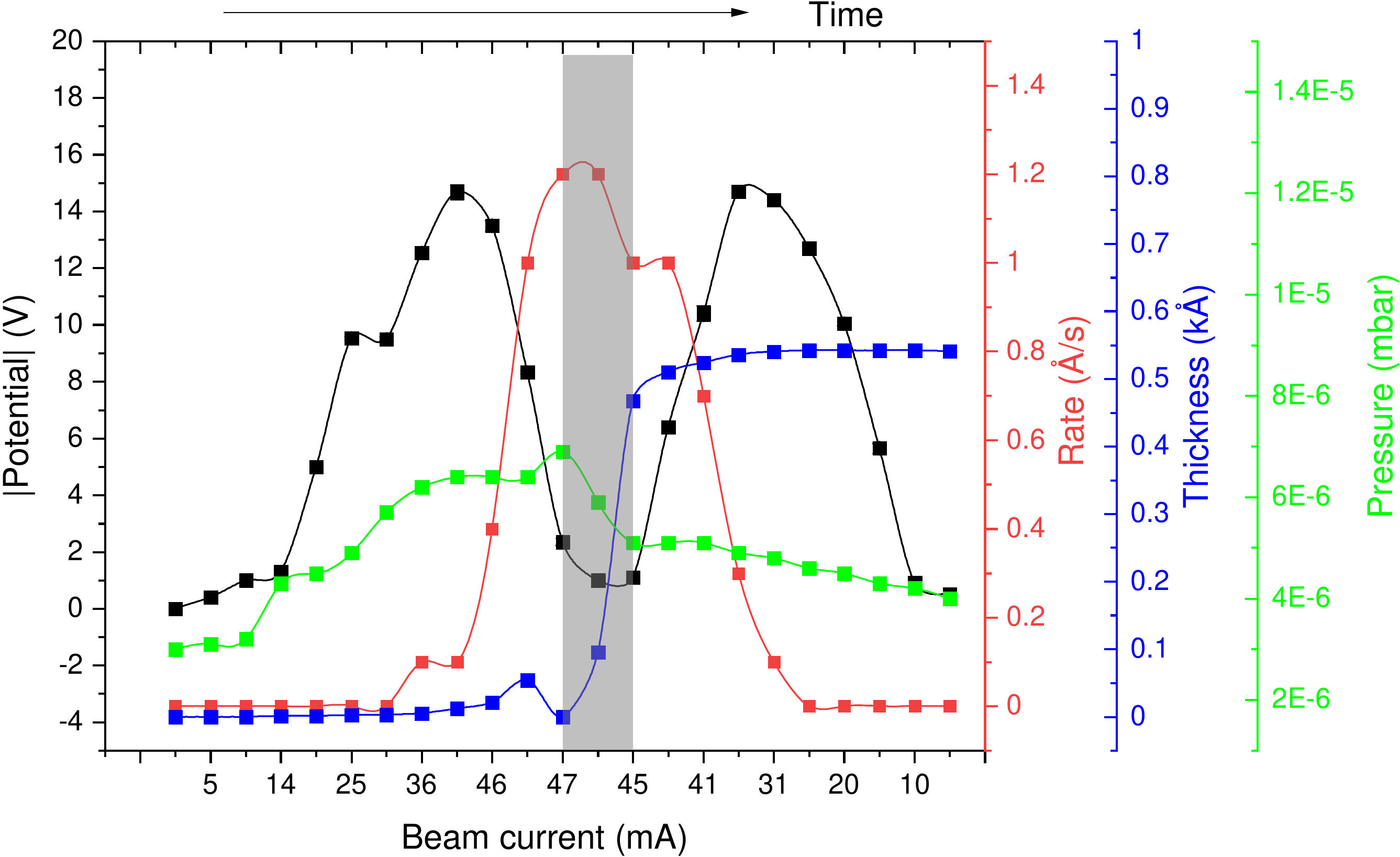}}
\caption{Potential at the cylindrical electrode, evaporation rate, film thickness (not corrected by tooling factor) and chamber pressure as a function of the electron beam current during Ni evaporation. Time window for an open shutter is indicated in the shaded area.}
\label{fig:Ni_cylinder_magnet}
\end{figure}
\begin{figure}[H]
\centerline{\includegraphics[width=\columnwidth]{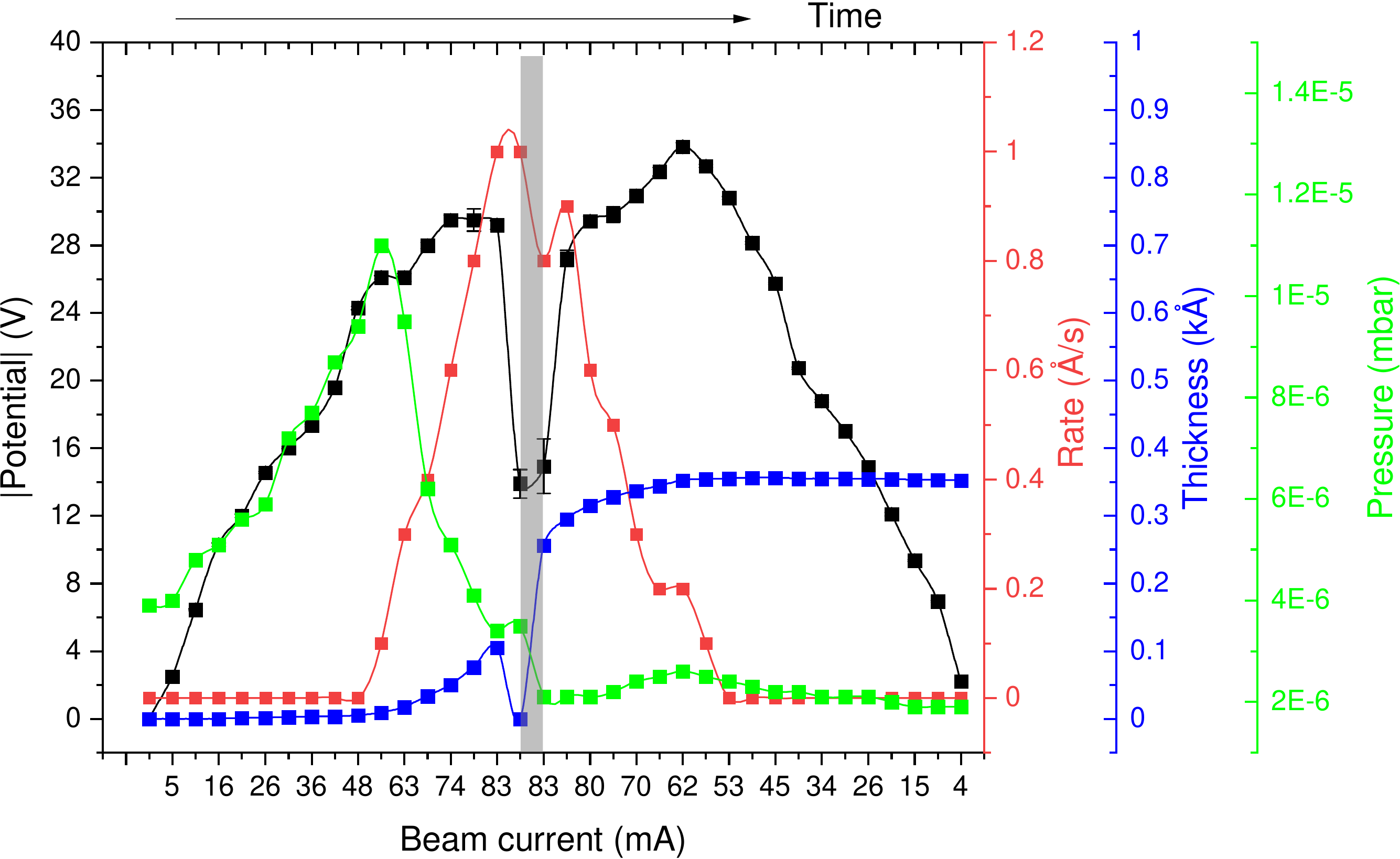}}
\caption{Potential at the cylindrical electrode, evaporation rate, film thickness (not corrected by tooling factor) and chamber pressure as a function of the electron beam current during Hf evaporation. Time window for an open shutter is indicated in the shaded area.}
\label{fig:Hf_cylinder_magnet}
\end{figure}
\begin{figure}[H]
\centerline{\includegraphics[width=\columnwidth]{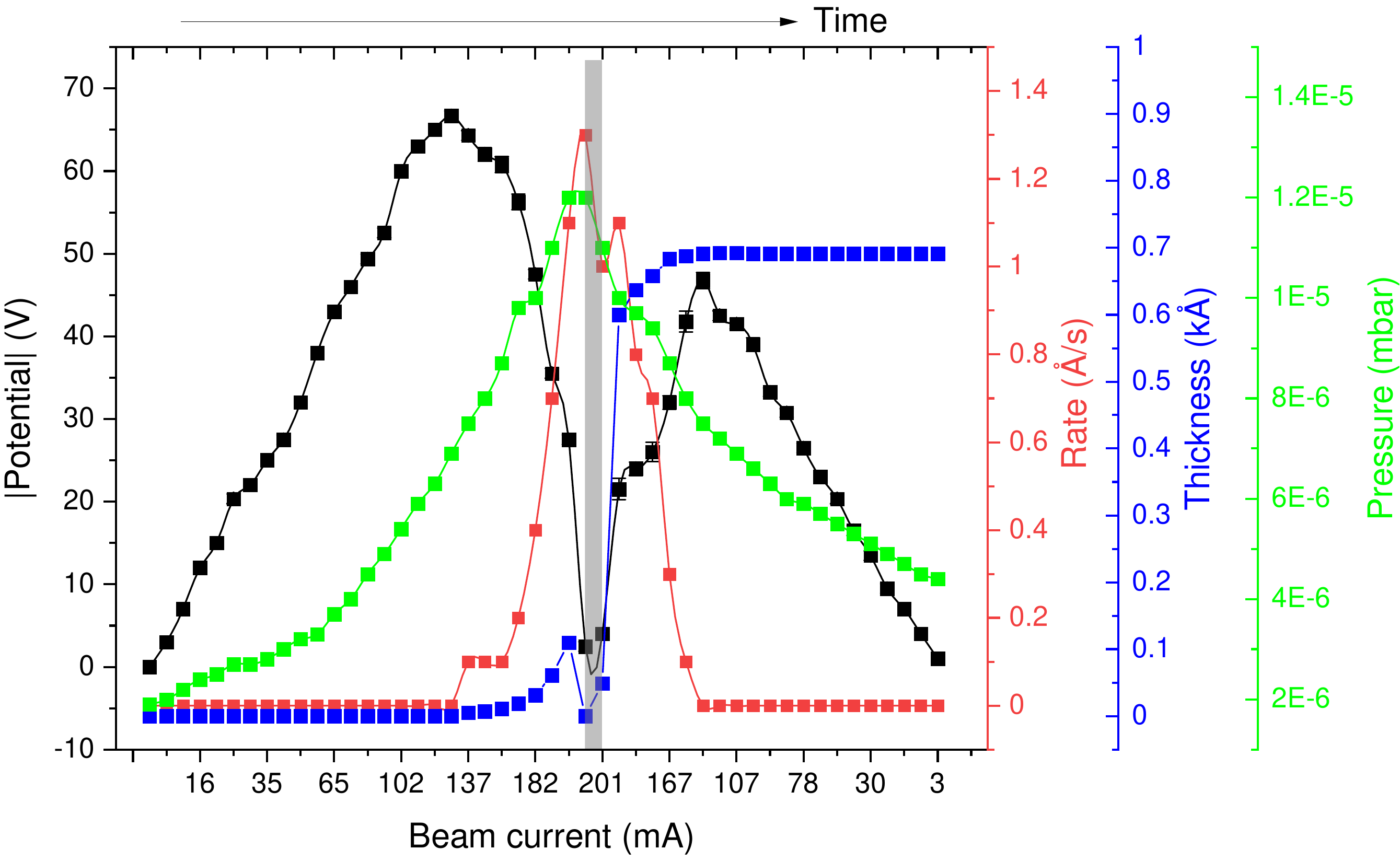}}
\caption{Potential at the cylindrical electrode, evaporation rate, film thickness (not corrected by tooling factor) and chamber pressure as a function of the electron beam current during Au evaporation. Time window for an open shutter is indicated in the shaded area.}
\label{fig:Au_cylinder_magnet}
\end{figure}

\section*{\label{app:subsec}Bibliography}
\nocite{*}
\bibliography{main}

\end{document}